# Controlling the micellar morphology of binary PEO-PCL block copolymers in water-THF through controlled blending

*Peter Schuetz[1*], Martin J. Greenall[2], Julian Bent[1], Steve Furzeland[1], Derek Atkins[1], Michael F. Butler[1], Tom C.B. McLeish[3], D. Martin A. Buzza[4]*

[1]Unilever R&D Colworth, Colworth Park, Sharnbrook, MK44 1LQ, UK

[2]School of Physics & Astronomy, University of Leeds, Leeds LS2 9JT, UK

[3]Department of Physics, Durham University, South Road, Durham DH1 3LE, UK

[4]Department of Physics, University of Hull, Hull HU6 7RX

*Corresponding author. e-mail: peter.schuetz@unilever.com

**Abstract:**

We study both experimentally and theoretically the self-assembly of binary polycaprolactone-polyethyleneoxide (PCL-PEO) block copolymers in dilute solution, where self-assembly is triggered by changing the solvent from the common good solvent THF to the selective solvent water, and where the two species on their own in water form vesicles and spherical micelles respectively. We find that in water the inter-micellar exchange of these block copolymers is extremely slow so that the resultant self-assembled structures are in local but not global equilibrium (i.e., they are non-ergodic). This opens up the possibility of controlling micelle morphology both thermodynamically and kinetically. Specifically, when the two species are first molecularly dissolved in THF before mixing and self-assembly ('pre-mixing') by dilution with water, the morphology of the formed structures is found to depend on the mixing ratio of the two species, going gradually on a route of decreasing surface curvature from vesicles *via* an intermediate regime of micelles in the shape of 'bulbed' rods, rings, Y-junctions finally to spherical micelles as we increase the proportion of the "sphere formers". On the other hand, if the two species are first partially self-assembled (by partial exchange of the solvent with water) before mixing and further self-assembly ('intermediate mixing'), novel metastable structures, including nanoscopic 'pouches', emerge. These experimental results are corroborated by self-consistent field theory calculations (SCFT) which reproduce the sequence of morphologies seen in the pre-mixing experiments. SCFT also reveals a clear coupling between polymer composition and aggregate curvature, with regions of positive and negative curvature being stabilized by an enrichment and depletion of sphere formers respectively. Our study demonstrates that both thermodynamic and kinetic blending of block copolymers are effective design parameters to control the resulting structures and allow us to access a much richer range of nano-morphologies than is possible with monomodal block copolymer solutions.

**1. Introduction**

The spontaneous formation of discrete structures (i.e. vesicles and micelles) from molecules with amphiphilic (surfactant) character provides a challenging and intriguing example of implicitly-controlled self-assembly.[1] These amphiphilic molecules can be "classic" molecular surfactants, or block co-polymers with differential attraction to the solvent, as in the case studied here. Block copolymer self-assembly is attractive for two reasons: Firstly as a route towards a fundamental understanding of self-assembly and emergent complexity it is compelling, since the underlying ingredients of polymer physics of local interactions and conformational entropy are well-understood.[2] For example, self-consistent field theory (SCFT) based on Gaussian chain entropy and local excluded volume has successfully mapped the fascinating emergent periodic structures that arise from local phase-separation in block co-polymer melts[3,4] as well as the range of structures obtained in diluted copolymer systems with different conformational asymmetry.[5,6] Secondly there is strong technological value to mastering the molecular control of block copolymer aggregates since they offer important materials advantages compared to self–assembled materials formed by low molecular weight amphiphiles. For example, block copolymer vesicles (i.e., polymersomes) are claimed to be more effective encapsulating vehicles compared to vesicles made from low-molecular weight surfactants due to their large internal volume and the low permeability and high tenability of their membrane walls.[7,8] Vesicles and micelles formed by block copolymers can also be cross-linked without structural disruption, leading to dramatic improvements in the structural stability of these aggregates.[8,9] Finally, block copolymers can form aggregates with novel topologies not seen for low molecular weight surfactants.[10]

The most obvious design parameter for self-assembled block copolymer systems is the fraction of both hydrophilic and hydrophobic components in the block copolymer chain.[7,8,10-13] For example, our previous results[11] showed that in a series of polycaprolactone-*block*-polyethyleneoxide (PCL-PEO) block-copolymers with various volume fractions of the hydrophilic block (PEO), high volume fractions $f_{EO}$ of the hydrophilic block (PEO) resulted in micelles ($f_{EO}$>0.3), whilst lower fractions favored wormlike micelles (0.25<$f_{EO}$<0.3) and finally vesicles ($f_{EO}$<0.25). This sequence of structures is in good agreement with the findings of previous authors[14] and was also recently replicated using real-space self consistent field theory (SCFT) modeling.[5,6,15] The self-assembly of block copolymers in solution into different morphologies has been explained qualitatively in terms of a balance between the steric repulsion of hydrophilic chains tethered at the interface (crowding) and the interfacial tension at the block junction, which increases with decreasing solvent quality for the hydrophobic block.[14] More recently however we have proposed a more rigorous theoretical model based Self-Consistent Field Theory that explains these changes quantitatively.[6]

In this study, we seek to control micellar morphology through controlled blending of block copolymers of different architectures. The use of blending as a control parameter for aggregate shape is well known in biological systems, for example local composition fluctuations of different lipid molecules in the membranes of cells can couple to membrane curvature to assist shape-driven processes such as budding and endosome production.[16] For block copolymer systems, Jain and Bates[17] have studied binary mixtures of poly(ethylene oxide)-poly(butadiene) (PEO-PB) block-copolymers in water and found that micelle morphology could be controlled by blending and that for specific blending conditions, intriguing morphologies not seen in monomodal systems could be produced, e.g., undulating cylinders and octopus-like structures.

In this paper, we study the micelle morphologies formed by binary mixtures of PEO-PCL, where the individual components form vesicles and spherical micelles respectively in water. Our study goes beyond that of Jain and Bates as in addition to studying the effect of mixing ratio on micelle morphology, we also study the effect of blending history on micelle morphology, specifically by allowing the two components to self-assemble to different degrees prior to mixing. This is interesting since one of the distinctive features of block copolymers compared to low molecular weight surfactants is their extremely slow inter-micellar exchange in a selective solvent (due to the very low critical micelle concentrations) so that the resultant aggregates tend to be non-ergodic.[17,18] This means that it should be possible to use blending history, in addition to blending ratio, as a control parameter for micelle morphology. In addition, we complement our experimental study by performing Self-Consistent Field Theory (SCFT) calculations of our binary block copolymer system. This allows us to understand quantitatively the relative stability of the different aggregates as well as the mechanism by which these aggregates are stabilized.

Recent literature describes at least three distinct methods to trigger the self-assembly of amphiphilic block copolymer systems: In the simplest case, only viable for sufficiently hydrophilic block copolymers, micelles can be formed by direct hydration from a surface film. The water diffuses into a pre-ordered lamellar block copolymer film, formed from a dried down solution in a good solvent, and the outer most layers bud off to form vesicles.[7-10,12,17-19] Another method forms ordered aggregates by self-assembly directly in solution. The block copolymer is initially molecularly dissolved in a common good solvent for both blocks. A subsequent change in solvent *condition* reduces the solubility of one block and triggers the self-assembly process. For suitable polymer blocks, this change in solubility can be induced by a change in temperature, ionic strength or pH.[20-22] However in this paper, we use a solvent switch method[11,14] where the block copolymers are first dissolved in a common solvent for both blocks (for our PEO-PCL polymers we use THF) and self-assembly is then triggered by mixing in a selective solvent (we use water as a selective solvent for PEO). If required, the organic solvent fraction in the mixture can be further reduced by a final dialysis step.

The main body of this paper is organized as follows: In section 2, we present the experimental details of our study, including both materials and methods. In section 3, we present results for the binary mixtures of PEO-PCL in THF-water, where the two components are blended before self-assembly ('pre-mixing'), after self-assembly ('post-mixing') and part way through self-assembly ('intermediate-mixing'). These systems were analysed experimentally using cryo-TEM, dynamic light scattering (DLS), NMR, turbidity measurements and theoretically using self-consistent field theory (SCFT). Finally in section 4, we present our conclusions.

## 2. Experimental Section

**Materials.** The PEO-PCL block copolymers were purchased from Advanced Polymer Materials Inc., Montreal and used as received. GPC analysis was provided by Advanced Polymer Materials Inc., referenced against PEO standards. Degrees of polymerisation for the PCL block were calculated by $^1$H NMR in CDCl$_3$ by comparison to the PEO block (the degrees of polymerisation for the monomethoxypoly(ethylene oxides) used in these polymerisations are known). The molecular weight and molecular weight distributions are given in Table 1. All other reagents with the exception of NMR solvents were purchased from Sigma Aldrich Company Ltd., Gillingham. Standard solvents were of spectrophotometric grade and inhibitor free. Deuterated NMR solvents were purchased from Euriso-top S.A., Saint-Aubin. All solvents were filtered before use through Pall Acrodisc PSF GHP 200 nm filters. For all experiments distilled, de-ionised Millipore water (resistivity = 18.2 MΩ.cm) was additionally filtered through Sartorius Ministart 200 nm filters directly before use.

| Commercial sample code | Sample formula | $M_w$[a] | $M_w/M_n$ | $f_{EO}$[b] | Morph-ology[c] | $R_h$[d] (nm) |
|---|---|---|---|---|---|---|
| PCL$_{5k}$PEO$_{1k}$ | PEO$_{23}$-b-PCL$_{47}$ | 7100 | 1.15 | 0.17 | V | 170 |
| PCL$_{5k}$PEO$_{2k}$ | PEO$_{45}$-b-PCL$_{43}$ | 7800 | 1.16 | 0.3 | S | 30 |

Table 1: Block copolymers used in this study. Notes: (a) total molecular weight from GPC (PEO standards) (b) volume fraction of the EO block calculated from the melt densities of the two blocks (c) morphology determined by light scattering and cryo-TEM (S=spherical micelles, C = worm-like micelle, V = vesicle; in the cases of mixed morphologies the majority component is written first), (d) hydrodynamic radius from dynamic light scattering (DLS) after dialysis.

**Preparation of Solutions.** Aqueous dispersions of block-copolymer aggregates were prepared by dissolving the polymer in THF to a concentration of 10 mg ml$^{-1}$. These solutions were then mixed in the volume ratio noted for the experiments. All the described mixing ratios are thus ratios of the masses of the respective polymers as opposed to molar ratios though molar ratios can easily be calculated since the molecular weights are known. In order to change the solvent quality for the PCL block, millipore water was added to the block copolymer solutions either manually or by an Eppendorf EDOS 5222 Electronic Dispensing System. 125 aliquots of 20 µl of water were added in one minute intervals until the desired THF fraction in water was obtained.

**Turbidity Measurements.** Preparation of samples with simultaneous turbidity measurements were performed using an adapted Perkin Elmer UV/Vis Lambda 40 Spectrometer. A wavelength of 600 nm was used with a slit width of 2 nm. Stirring was achieved using a standard magnetic stirrer / hotplate placed under the spectrometer. The polymer was dissolved in THF (1 mL, 10 mg mL$^{-1}$) and a zero reading was taken (transmittance, T = 100%). Millipore water was then added either in 10 µl aliquots every 30 s using an Eppendorf EDOS 5222 Electronic Dispensing System with turbidity reading taken after each addition

**Cryo-TEM.** Samples for thin-film cryo-TEM were loaded onto plasma treated (30 seconds) holey-carbon grids and prepared using a GATAN cryo-plunge into liquid ethane and then transferred using a GATAN 626 cryo-transfer system. Samples were examined using a JEOL 2100 TEM operating at 200kV. Images were obtained using a Bioscan or a GATAN Ultrascan 4k camera and analysed by GATAN Digital Micrograph version 1.71.38. During our previous investigations we observed that imaging the self-assembled structures (especially vesicles) by cryo-TEM is greatly improved in samples containing ca. 30% THF compared to samples in pure aqueous solution.[11] This is probably due to a combination of different wetting conditions of the solutions on the carbon film as well as a higher membrane flexibility of the vesicles. Dynamic light scattering results showed that the diameters of vesicles in 28% THF were substantially larger than to those in water due to de-swelling of the structures as THF is removed. This indicates that at THF fractions in water of 30% and below, the mobility of the block-copolymer chains is too restricted to allow for further shape changes of the aggregates. All cryo-TEM images in this paper are therefore reported for THF fractions in water of 28%.

**Dynamic light scattering (DLS)** measurements were carried out on a Malvern Instruments Nano-ZS Zetasizer. The measurements were made at a scattering angle θ=173° at 25 °C. Data were acquired using Malvern Instruments' Dispersion Technology Software (DTS) version 4.20. The autocorrelation functions were analysed with the cumulants method to obtain the translational diffusion coefficient. The software package of the instrument calculates the hydrodynamic radius and the polydispersity of the sample using the Stoke-Einstein equation. Appropriately corrected viscosity and refractive index values were used for each solvent mixing ratio.

**Diffusion Ordered Spectroscopy (DOSY-NMR)** measurements were conducted using a bbi probe and a Bruker AV600 spectrometer. Solution state $^1$H spectra were acquired using the standard ledbpgppr2s Bruker pulse program, with a diffusion time of (Δ) 150ms, a gradient pulse length (δ) of 15ms and increasing the gradient strength $5 < g < 32$ Gcm$^{-1}$. The water resonance from the solvent (28% THF-d8 72% H$_2$O) was suppressed with low power, continuous wave gradients during the relaxation delay whilst the THF resonance was minimized by using deuterated THF. The proton frequency was locked relative to D$_2$O and the chemical shift referenced relative to TMS, both in a capillary. The self diffusion (D) of the block co-polymer aggregates was calculated using the attenuated intensity of a PCL peaks between $1.75 > $ ppm $ > 1$, according to the Stejskal-Tanner equation

$$\frac{I}{I_0} = \exp\left[-\gamma^2 g^2 \delta^2 D(\Delta - \delta/3)\right]$$

where γ is the gyromagetic ratio (4257HzG$^{-1}$). The self diffusion coefficient was then used to calculate the radius of hydration ($R_H$) using the Stokes Einstein relation

$$R_H = \frac{k_B T}{6\pi D}$$

where $k_B$ is Boltzmann's constant ($1.38 \times 10^{23} \text{JK}^{-1}$) $T$ is the absolute temperature. 1d $^1$H NMR spectra were also acquired, using the standard noesygppr1d Bruker pulse program. The intensities and chemical shift were normalized using the TMS resonance at 0 ppm.

## 3. Results

**Pre-mixed vs. Post-mixed Experiments**

As can be seen from Table 1 and Figure 1 a,b, the two block-copolymers PCL$_{5k}$PEO$_{1k}$ and PCL$_{5k}$PEO$_{2k}$ form vesicles and spherical micelles respectively, when assembled on their own by solvent exchange from THF. By mixing the two block copolymers before and after their self assembly (i.e. in solutions 100% THF and 28% THF in water, respectively), we can estimate the timescale for inter-micellar exchange of block copolymer chains.[10,18,23] To this end, in one case, we dissolved the two polymers individually in THF, mixed the PCL$_{5k}$PEO$_{1k}$ and PCL$_{5k}$PEO$_{2k}$ solutions in a 3:1 ratio, then induced self-assembly in the mixture by solvent exchange to 28% THF in water; we call this protocol 'pre-mixing'. In another case, we dissolved the two polymers individually in THF, induced self-assembly in each solution by solvent exchange to 28% THF in water, then mixed the PCL$_{5k}$PEO$_{1k}$ and PCL$_{5k}$PEO$_{2k}$ solutions in a 3:1 ratio; we call this protocol 'post-mixing'. In Figure 1c, we see that in the pre-mixing experiment, large vesicles with diameters around 400 to 500 nm are formed, similar to the ones seen for PCL$_{5k}$PEO$_{1k}$ alone. However in Figure 1d, we see that in the 'post-mixing' experiment, vesicle and spherical micelle morphologies coexist with each other even after ageing at room temperature for a few weeks. These results indicate that inter-micellar exchange of block copolymer chains in 28% THF in water is negligible on the timescale of weeks. These results are in good agreement with similar experiments by Jain and Bates for binary mixtures of PEO-PB in water.[17]

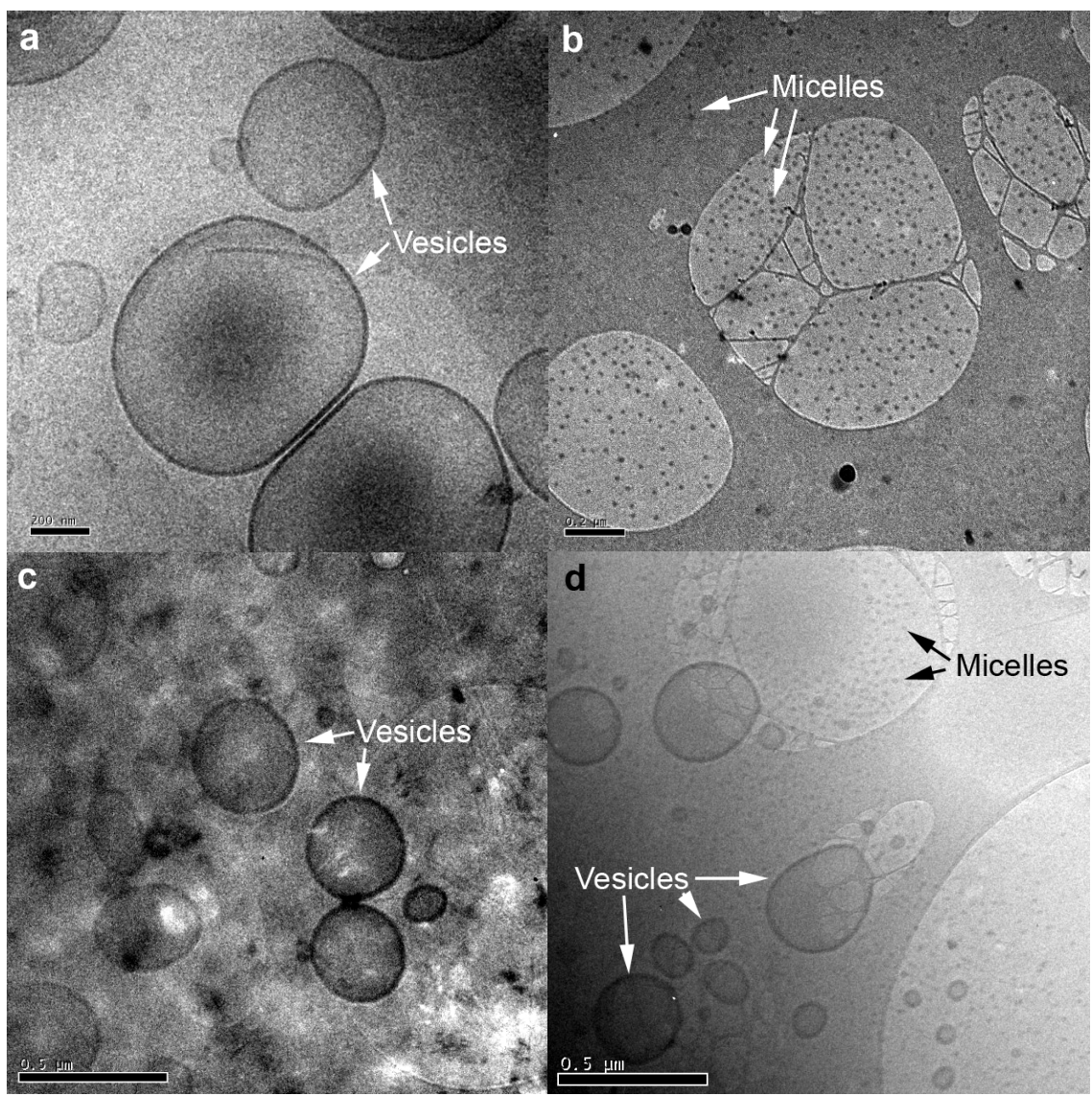

Figure 1: Cryo-TEM images of self-assembled structures from $PCL_{5k}PEO_{1k}$ and $PCL_{5k}PEO_{2k}$ mixtures in 28% THF. (a) pure $PCL_{5k}PEO_{1k}$ (b) pure $PCL_{5k}PEO_{2k}$ (c) $PCL_{5k}PEO_{1k}$ and $PCL_{5k}PEO_{2k}$ mixed in a ratio 3:1 prior to self-assembly ('pre-mixed' sample) (d) $PCL_{5k}PEO_{1k}$ and $PCL_{5k}PEO_{2k}$ mixed in a ratio 3:1 after self-assembly ('post-mixed' sample). All images have the same magnification; the scalebars are 200 nm for the top images and 500 nm for the ones at the bottom.

However, it should be noted however that, although the cryo-TEM results are striking and very informative, the liquid film on the carbon grid is 100s of nm thick before vitrification,[24,25] and therefore the vesicles within it are subject to confinement effects during blotting (removal of excess water). This can be seen from the fact that in the cryo-TEM images, the vesicles generally prefer to sit in the hole areas of the carbon grids (see for example Figure 6) or, alternatively, close to the grid meshes, as these are the areas where after blotting the thickest liquid films remain. In order to confirm the cryo-TEM results above and correct for any confinement effects, we therefore also measured aggregate sizes in both the pre-mixed and post-mixed solutions using the bulk methods of dynamic light scattering (DLS) and DOSY-NMR.

From Table 2, we see that DLS confirms the presence of vesicles with diameters around 400 to 800 nm in both the pre-mixed and post-mixed samples. Unfortunately, the spherical micelles which are clearly seen in the cryo-TEM image of the post-mixed solution could not be detected by DLS as the signal is dominated by the strongly scattering vesicles obscuring any spherical micelles present (these are more than one order of magnitude smaller and hence scatter more than three orders of magnitude less for similar phase volumes).

|  |  | DOSY-NMR | | DLS | |
|---|---|---|---|---|---|
|  |  | $D\ (m^2 s^{-1})$ | $R_h$ (nm) | $D\ (m^2 s^{-1})$ | $R_h$ (nm) |
| $PCL_{5k}PEO_{1k}$ | vesicles | $4.4 \times 10^{-12}$ | 64 | $5.20 \times 10^{-13}$ | 500 |
| $PCL_{5k}PEO_{2k}$ | micelles | $1.1 \times 10^{-11}$ | 26 | $1.05 \times 10^{-11}$ | 28 |
| Pre-mixed | vesicles | $4.2 \times 10^{-12}$ | 68 | $4.6 \times 10^{-13}$ | 472 |
| Post-mixed (fast) | micelles | $1.4 \times 10^{-11}$ | 20 | n/a | n/a |
| (slow) | vesicles | $3.6 \times 10^{-12}$ | 79 | $1.3 \times 10^{-12}$ | 226 |

Table 2: Diffusion coefficients and corresponding hydrodynamic radii from DOSY-NMR and DLS for the self-assembled structures of $PCL_{5k}PEO_{1k}$ and $PCL_{5k}PEO_{2k}$ as well as pre- and post-mixed samples.

For this reason, the presence of vesicles and/or spherical micelles was also be measured by diffusion-ordered NMR spectroscopy (DOSY-NMR). This technique resolves solution state NMR spectra in a second dimension, ordered by diffusion coefficient. As this measure of diffusion is intrinsically weighted by the number average of protons ($^1$H-NMR spectra are recorded with increasing pulsed field gradient strength) it is not skewed by aggregate size, unlike in light scattering techniques. The decay of the PCL resonances between 1.7 > ppm > 1 for each sample are shown in Figure 2; the PEO resonance at 3.7 ppm was not used for this analysis as it was obscured by the THF (solvent) resonance. The diffusion coefficients, summarized in Table 2, were obtained by fitting with one or two exponential functions allowing us to characterize the size of aggregates in mixtures containing either one or two types of aggregates.

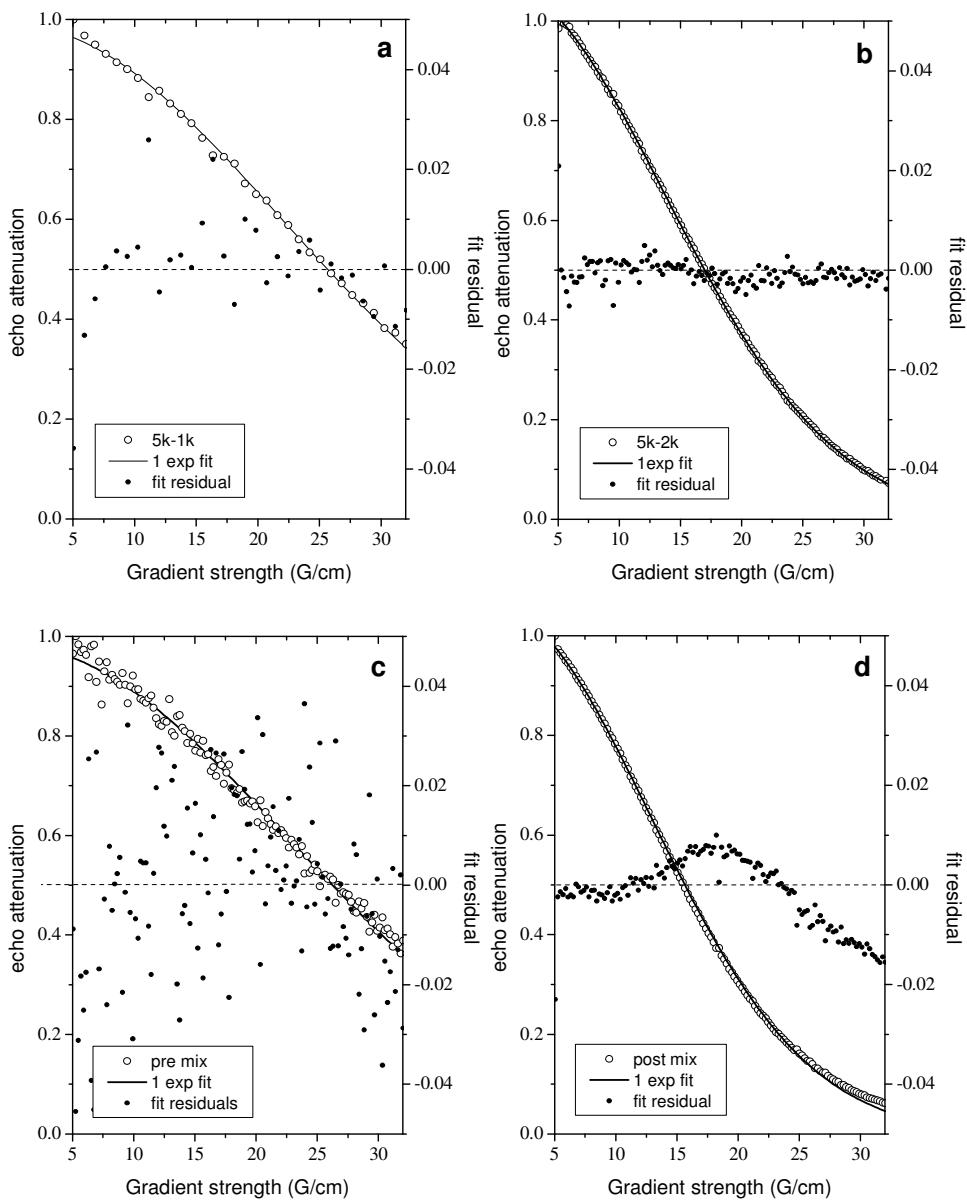

Figure 2: Single exponential fits (and residuals) to the PCL peak attenuation with increasing gradient strength in (a) $PCL_{5k}PEO_{1k}$, (b) $PCL_{5k}PEO_{2k}$, (c) pre-mixed $PCL_{5k}PEO_{1k}$ and $PCL_{5k}PEO_{2k}$ (d) post mixed $PCL_{5k}PEO_{1k}$ and $PCL_{5k}PEO_{2k}$, all in 28%THF-d8 and 72% water.

Table 2 lists the diffusion coefficients measured by DOSY-NMR and corresponding, calculated hydrodynamic radii ($R_H$) of the self-assembled structures of $PCL_{5k}PEO_{1k}$ (vesicles), $PCL_{5k}PEO_{2k}$ (spherical micelles) and pre-mixed and post-mixed solutions of these polymers, all dissolved in 28% THF. The $PCL_{5k}$-$PEO_{1k}$ and the pre-mixed solution DOSY-NMR data both decay with a slow diffusion coefficient, corresponding to a large assemblies (i.e. vesicles), whilst the $PCl_{5k}$-$PEO_{2k}$ and the post-mixed sample data decay faster corresponding to smaller assemblies (i.e. micelles). Furthermore, the post-mixed sample data is not best described by a single exponential but by a weighted sum of the micelle and vesicle components. This is in good agreement with the cryo-TEM images in Figure 1 and complements the DLS measurements which could not detect the presence of spherical micelles in the post-mixed sample.

However we note that while the sizes of the spherical micelles measured by DOSY-NMR and DLS are in good agreement with each other, the signal intensities (indicated here by signal to noise ratio) are weaker for the slow diffusing component and the corresponding size of the vesicles measured by DOSY-NMR is underestimated by almost an order of magnitude compared to DLS. We believe that this due to the fact that the PCL domains in vesicles are more crystalline than those in spherical micelles. This is confirmed by an increase in intensity and narrowing of PCL $^1$H NMR lineshape with increasing temperature. The relaxation of the NMR resonances is not only due to the diffusion of aggregates, but is also dependant on the intrinsic spin relaxation rate of the exited nuclei, which is faster in a more rigid, crystalline state. This means that not all the vesicle protons are measured in these experiments and hence the sizes are not as accurate as when measured by DLS and TEM. Notwithstanding this fact, DOSY-NMR clearly identifies the presence of both vesicles and spherical micelles in the post-mixed sample, in good agreement with the cryo-TEM results.

**Pre-mixed Experiments with Variable Mixing Ratio**

In the previous section, we described micelle morphologies obtained from mixing vesicle formers (PCL$_{5k}$PEO$_{1k}$) to sphere formers (PCL$_{5k}$PEO$_{2k}$) at a specific ratio of 3:1. In this section, we study the effect of changing the mixing ratio of the two polymers on the micelle morphologies obtained. To ensure maximal mixing during structure formation, all PCL$_{5k}$PEO$_{1k}$ and PCL$_{5k}$PEO$_{2k}$ mixtures in this section were thus prepared using the pre-mixing protocol where both polymers were first individually dissolved in THF and mixed to the desired stoichiometry before the self-assembly was initiated by dilution with water down to 28% THF (i.e., 72% water). Later in the paper, we allow the individual polymers to partially self-assemble before mixing in order to study the effect of blending history on micelle morphology. To observe the effect of solvent quality on structure formation, we used turbidity traces during dilution (i.e., transmission as a function of water content). The final solutions containing 28% THF were then characterized by DLS and, for selected samples, by cryo-TEM.[23]

Figure 3 shows selected turbidity traces from a series of different mixing ratios of $PCL_{5k}PEO_{1k}$ and $PCL_{5k}PEO_{2k}$. As discussed in our previous paper, the prominent trough in the transmission at around 30 to 35% water content that is seen for all polymer mixing ratios is most probably due to a miscibility gap in the PEO-THF-water phase diagram.[11,26,27] On increasing the $PCL_{5k}PEO_{2k}$ fraction, the dip in transmission around 40% to 50% water content, which is linked to the formation of wormlike micelles, is gradually reduced and finally vanishes for compositions containing around 40% $PCL_{5k}PEO_{2k}$. For $PCL_{5k}PEO_{2k}$ compositions between 45% and 75%, the final transmission (i.e., the transmission at 72% water content) increases until at around 75% $PCL_{5k}PEO_{2k}$ the final solution is clear as in the pure $PCL_{5k}PEO_{2k}$ system.

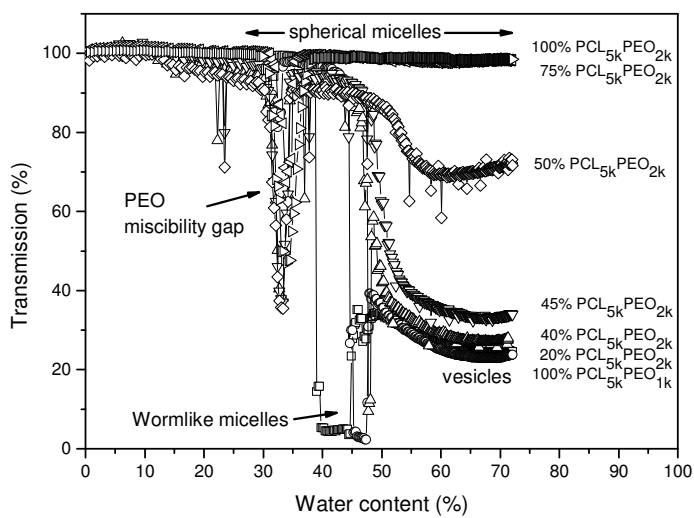

Figure 3: Turbidity traces for the self-assembly of block copolymer mixtures of $PCL_{5k}PEO_{1k}$ and $PCL_{5k}PEO_{2k}$. The optical transmission (in %) at 600 nm is plotted as a function of the water content of the solution.

The evaluation of the full data-set in Figure 4 highlights again these two trends. In Figure 4a, the water content of the onset of the transmission drop associated with wormlike micelles is plotted as function of $PCL_{5k}PEO_{2k}$ fraction and we see that the water content for the onset is increased as we increase the fraction of sphere former $PCL_{5k}PEO_{2k}$. In Figure 4b, the final transmission (at 72% water content) is plotted as a function of $PCL_{5k}PEO_{2k}$ fraction and we see that there is a sharp increase in the final transmission (around 50% $PCL_{5k}PEO_{2k}$) as we increase the fraction of $PCL_{5k}PEO_{2k}$. The results in Figure 4 point to dramatic changes in micelle morphology as we increase the fraction of sphere former $PCL_{5k}PEO_{2k}$.

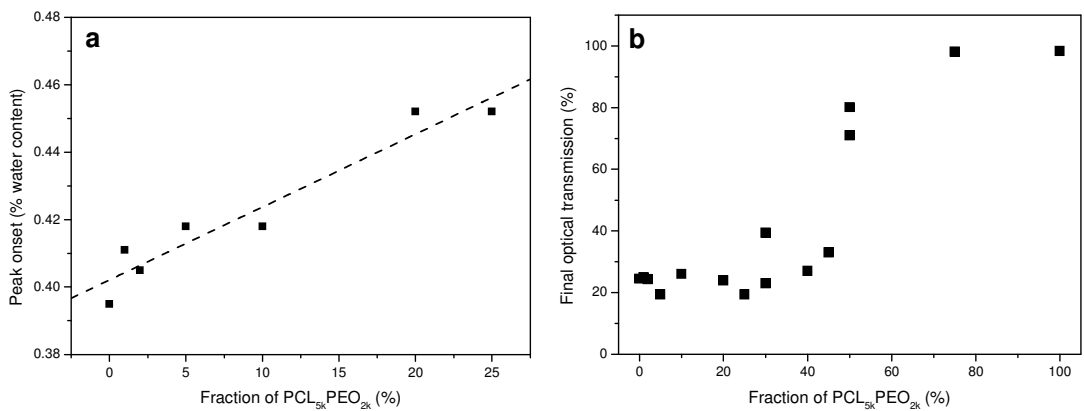

Figure 4: Data extracted from the turbidity traces of the mixed $PCL_{5k}PEO_{1k}$ and $PCL_{5k}PEO_{2k}$ systems in Figure . (a) water content for the onset of the transmission drop associated with wormlike micelles as function of $PCL_{5k}PEO_{2k}$ fraction (the dashed line is a guide to the eye) (b) the final transmission (at 72% water content) as a function of $PCL_{5k}PEO_{2k}$ fraction.

To investigate these changes in more detail, in Figure 5, we plot the hydrodynamic radii $R_H$ measured by DLS as a function of the fraction of $PCL_{5k}PEO_{2k}$ for the final solutions (i.e., containing 72% water). For $PCL_{5k}PEO_{2k}$ fractions below 30% where the solutions are turbid, the average aggregate size is around 950 nm, indicating the presence of vesicles (after dialysis, the average aggregate size for these solutions shrinks to about 450 nm). However for $PCL_{5k}PEO_{2k}$ fractions above 30%, the average aggregate size starts to fall until for $PCL_{5k}PEO_{2k}$ fractions above 75% where the solutions are clear, the average aggregate size plateaus to around 60 nm, indicating the presence of spherical micelles.

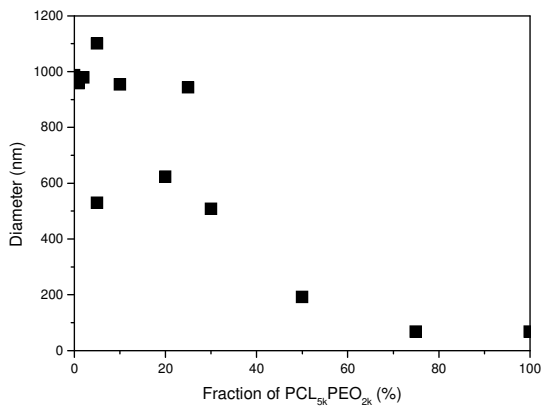

Figure 5: Hydrodynamic radii from dynamic light scattering (cumulants fit) as a function of $PCL_{5k}PEO_{2k}$ content for the self-assembled structures in mixtures of $PCL_{5k}PEO_{1k}$ and $PCL_{5k}PEO_{2k}$ in 28% THF.

Cryo-TEM images of the corresponding final solutions (i.e., 72% water) containing 5%, 25%, 50%, 75% of $PCL_{5k}PEO_{2k}$ are presented in Figure 6a,b,c,d, respectively. For both 5% and 25% of $PCL_{5k}PEO_{2k}$, where the final solutions are turbid, only vesicles are present, similar to solutions of pure $PCL_{5k}PEO_{1k}$ (shown above). For 50% of $PCL_{5k}PEO_{2k}$ in the mixture, where the final solution is much clearer than for lower $PCL_{5k}PEO_{2k}$ concentrations, a rich variety of morphologies is observed, including wormlike micelles, small vesicles (50 to 200 nm in diameter), short bulbous rods, small rings (similar in size to the vesicles) and also a small number of Y-junctions. Finally at 75% of $PCL_{5k}PEO_{2k}$, where the solution is clear, only spherical micelles are seen, similar to solutions of pure $PCL_{5k}PEO_{2k}$.

Interestingly, the morphology of the 50% binary blend is far richer than the morphology of a monomodal solution with the same average volume fraction of PEO ($f_{EO}$ = 0.235), where the latter forms vesicles only.[11] This is consistent with the results of Jain and Bates[17] who found that for certain mixing ratios, binary blends can form intriguing new morphologies not seen in their monomodal counterparts. As will be discussed in more detail below, our SCFT calculations indicate that the prevalence of a wide range of finite structures at intermediate mixing ratios for the binary system may be due to the coupling of polymer composition with aggregate curvature; this effect helps to stabilize edges that would otherwise be unstable. The coexistence of such a plethora of structures at 50% $PCL_{5k}PEO_{2k}$ also provides another clear indication that these systems are in local but not global equilibrium.

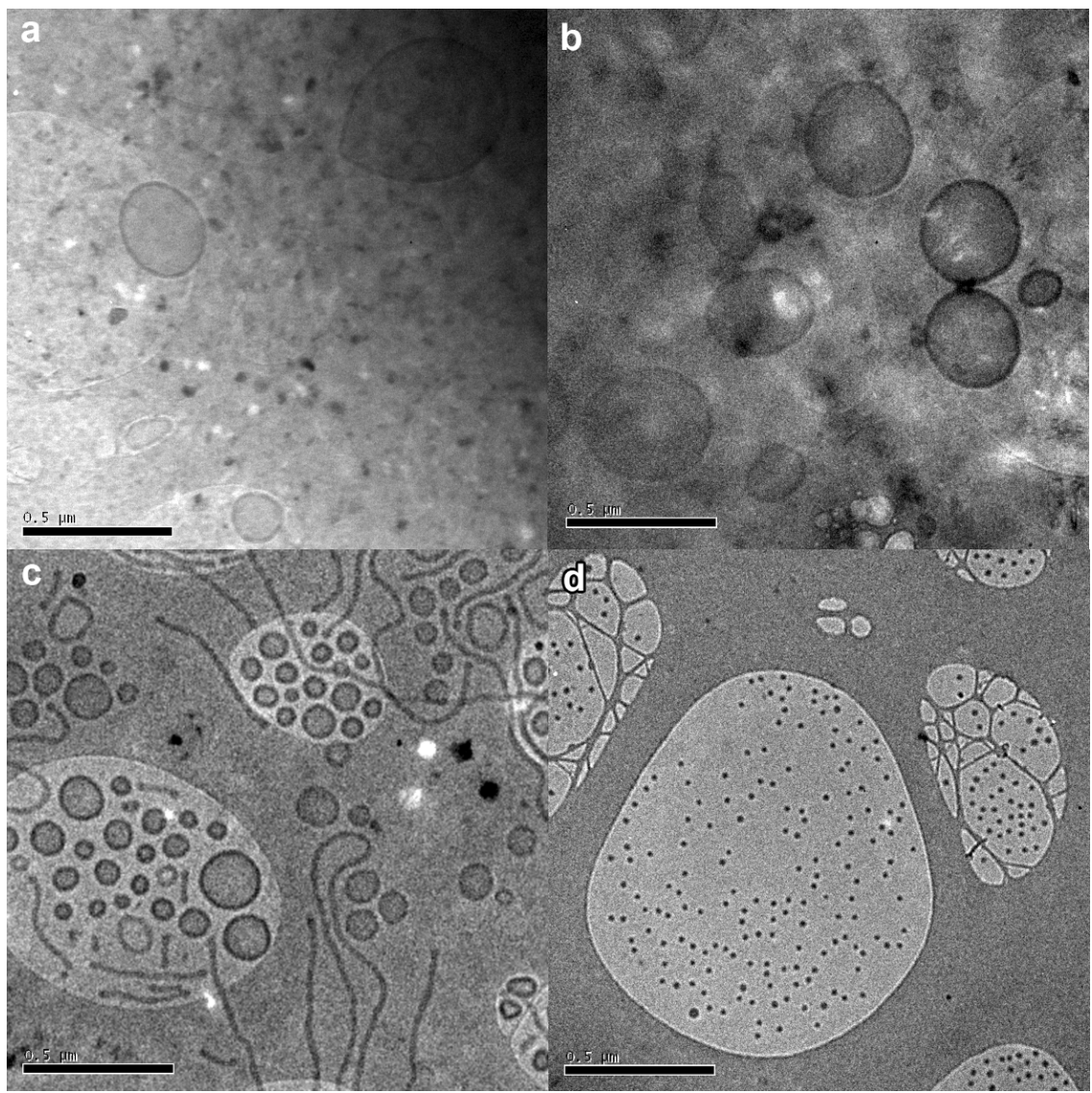

Figure 6: Cryo-TEM images PCL$_{5k}$PEO$_{1k}$ and PCL$_{5k}$PEO$_{2k}$ mixtures in 28% THF containing (a) 5% (b) 25% (c) 50% (d) 75% PCl$_{5k}$PEO$_{2k}$. The images at 5% and 25% PCL$_{5k}$PEO$_{2k}$ contain artifacts due to ice formation on the grids. All scalebars are 500 nm.

These cryo-TEM results confirm the DLS and turbidity results and show in much greater detail the strong effect that the addition of sphere former has on the micelle morphology of our binary blend. The aggregate sizes observed in these images also correlate well with the sizes measured DLS though the vesicle diameters observed in the cryo-TEM images are systematically smaller than the light scattering results. The latter point was also found in previous work[9,11,28,29] and is attributed to the confinement effects during sample preparation since the TEM films are only 100s of nm thick.[11]

**SCFT Calculations of Binary System**

In order to gain a better understanding of the complex self-assembly behavior exhibited by our binary block copolymer system, we performed self-consistent field theory (SCFT) calculations on a simple model system consisting of two types of AB diblock copolymers (lamella- and sphere-forming respectively) blended with A homopolymer 'solvent'. As in the experiments, the hydrophobic B-blocks of the two species of diblock are matched to contain the same number of monomers. The architectures of the two copolymer species were then chosen so that they preferentially form lamellae and spherical micelles respectively: the lamella-formers contain 50% hydrophilic A-blocks by volume, whilst the sphere-formers contain 75% A-blocks. The A homopolymer molecules ('solvent') were also taken to have the same length as the lamellar formers. We used a simple implementation of SCFT in which the individual polymer molecules are modeled by random walks and the interactions of the polymers are included by imposing incompressibility and introducing a contact potential between the A and B monomers, the strength of which is set by the Flory $\chi$ parameter.[2] This takes the moderate value of $\chi N = 30$, where $N$ is the number of monomers in the sphere-forming species. Although considerably simpler than the experimental system, this model contains enough detail to reproduce much of the experimental phenomenology.

Using SCFT, the free energy and density profiles were calculated for different morphologies, including spheres, rods, rings, vesicles and disks (platelets), at different mixing ratios of sphere and lamellar formers. For rods, rings and disks, we considered a single micelle in a *cylindrical* box while for vesicles and spherical micelles, we considered a single structure in a *spherical* box. In all cases, the SCFT diffusion equations were solved in real space using a standard finite-difference method[30] with spatial step size of 0.04 (working in length units where $N^{1/2}a = 1$, where $N$ is the number of monomers in the sphere former and $a$ is the monomer length for both A and B monomers) and imposing reflecting boundary conditions at both the origin and boundary of the box. Further details of our SCFT calculation can be found in refs.[6,15].

To make direct comparisons between the different structures, we carried out all calculations at the same block copolymer volume fraction (8%) and in equal-sized boxes. The exception was the case of spherical micelles, where we used the same block copolymer volume fraction but chose a box volume that minimized the free energy density of the system; as shown in a previous paper[15], this is equivalent to a system containing many micelles minimizing its free energy by adjusting the number (and hence the size) of the micelles. The reason for treating spherical micelles on a slightly different footing from the other structures is because while a non-spherical micelle can minimize its free energy per chain by changing its long dimension (rod length, disk radius etc.), such a minimization route is not available to spherical micelles[17]. Because of this, the free energy per chain for spherical micelles depends very strongly on aggregation number, unlike for all the other structures.[1] In our calculations for spherical micelles it is therefore important to use the equilibrium aggregation number or, equivalently, since we are working at fixed copolymer concentration, the equilibrium volume per micelle. However in order to make a direct comparison of our spherical micelle results with the other structures, we have normalized the free energy of the spherical micelle system by calculating the free energy of a box of the same size as the other structures containing the equilibrium number of spherical micelles.

Figure 7 presents the free-energy for all the different micelle structures (relative to that of the vesicle) as a function of the fraction of sphere former. The plot shows the transition of the lowest free energy morphology from lamellar vesicles to rings or rods (the free energies of these structures track each other very closely in SCFT) and eventually to spherical micelles. This is the same sequence as what is seen in the experiments, although, as expected from the necessary calculational use of a polymeric (rather than monomeric) solvent, the volume fractions at which transitions occur are slightly renormalized in our SCFT calculations. Specifically, theory predicts vesicles to be stable between 0% to 10% while in the pre-mixed experiments, vesicles were seen at a volume fraction of sphere former of 25% (Figure 6b). Theory predicts rods and rings to coexist between 10% to 50% while in the pre-mixed experiments, a coexistence of rings, rods and vesicles was seen at 50% sphere former (Figure 6c). We note that between 10% to 25% sphere former, our SCFT calculations find that the vesicle free energy is close to that of rings and rods. Recalling that our block copolymer system is non-ergodic, the proximity in the free energy between these different structures explains why a wide range of structures co-exist for this intermediate range of sphere formers. (A low density of Y-junctions was also observed in the 50% experiments though unfortunately these could not be modelled by our 2D SCFT calculations because they break cylindrical symmetry.) Finally theory predicts spheres to exist above 50% while in the experiments, spherical micelles were seen at a 75% level of sphere former (Figure 6d).

Interestingly, although the platelet structures did emerge as local solutions to the SCFT equations in the calculation, they never assumed the status of being the lowest free energy structure, nor even the next lowest free energy structure, regardless of the fraction of sphere former. This is again in agreement with the experiments where we did not observe any free platelet structures.

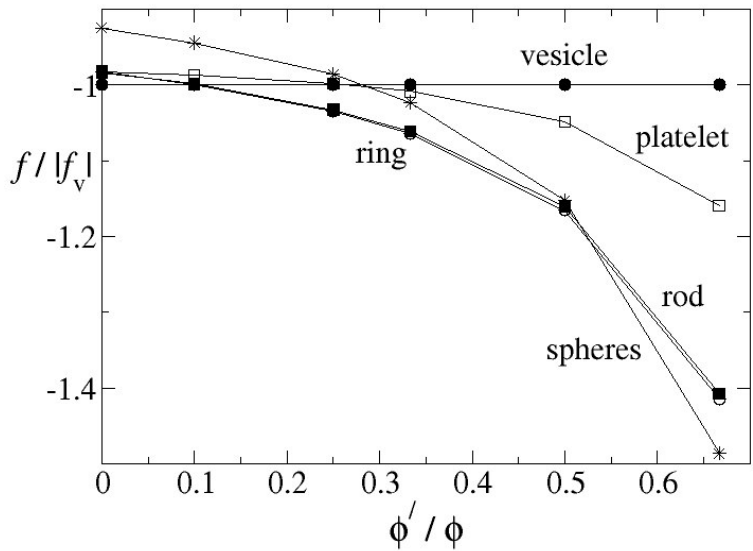

Figure 7: Plot of free energy density of each morphology (relative to that of the vesicle) against the fraction of sphere forming copolymer predicted by SCFT.

The SCFT calculations can be further interrogated to explore the mechanism by which the metastable structures of rods and platelets are stabilised. In particular we are interested in local variations in the concentration of the sphere-forming and lamella-forming polymers within the aggregates. For this purpose, we define an *enhancement factor* $\eta$ as the local ratio of the volume fraction of hydrophobic blocks from sphere formers to lamellar formers. In Figure 8, we plot the enhancement factor spatially as a function of the cylindrical co-ordinates for rods (a) and plates (b) within the hydrophobic core region, defined as the region where the volume fraction of hydrophobic blocks exceeds hydrophilic blocks. The concentration of the sphere-former towards the more highly-curved regions at the end of the rod or edge of the plate is immediately apparent.

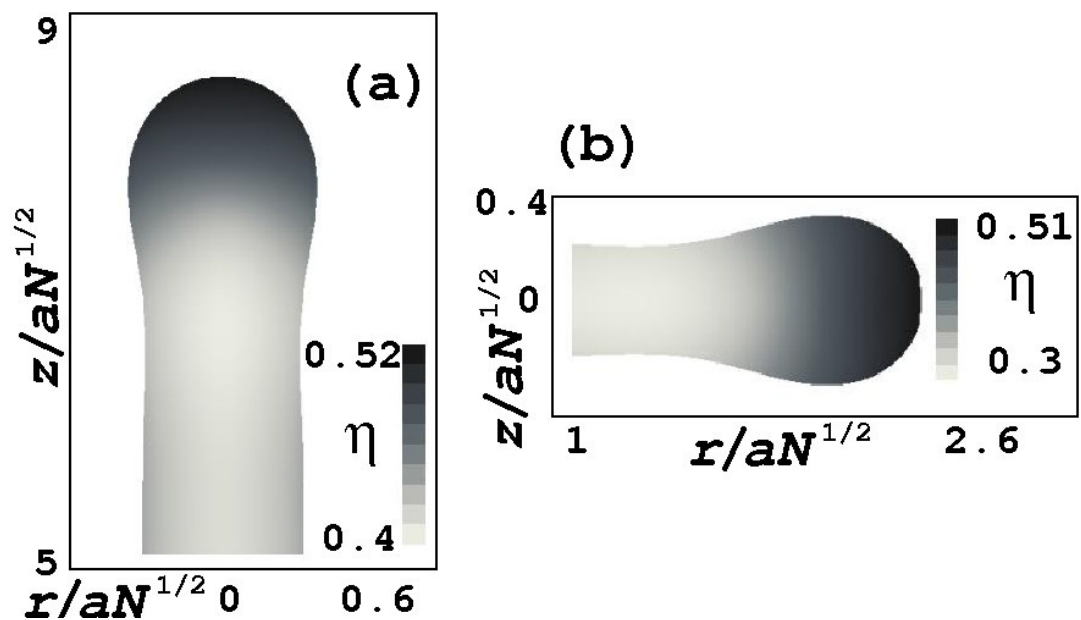

Figure 8: Plots in cylindrical polar coordinates of the enhancement factor $\eta$ for sphere formers in the hydrophobic core region (where $\eta$ is defined as the ratio of the local volume fraction of hydrophobic blocks from sphere formers to lamellar formers) from our SCFT calculations of (a) rods (b) platelets. The higher the value of $\eta$, the darker the shading.

Intriguingly, both structures also possess a small 'neck' region of *negative* curvature immediately prior to their terminator structure, giving the latter a bulbous appearance. The bulbous ends for both structures is also evident in Figure 9a,b where we plot the local density of the hydrophilic component of the block copolymers in the case of rods and platelets respectively in the cylindrical co-ordinates. In the case of rods, the bulbous end is also clearly seen in the microscopy of figure 6c. The negative curvature neck region in both structures can be attributed to depletion in the local concentration of sphere-formers as compared to the mean concentration far from the end-caps. This is most clearly seen in Figure 10 where we plot cuts through the volume fraction profiles of the different diblock species for rods (a) and plates (b), with the sphere forming species highlighted by heavier lines in both cases. From the volume fraction profile of the hydrophobic blocks (solid lines), we clearly see that for both rods and plates, the bulbous end is accompanied by an enhancement of sphere former while the negative curvature neck region is accompanied by a slight reduction in the local concentration of the sphere-former relative to lamellar former.

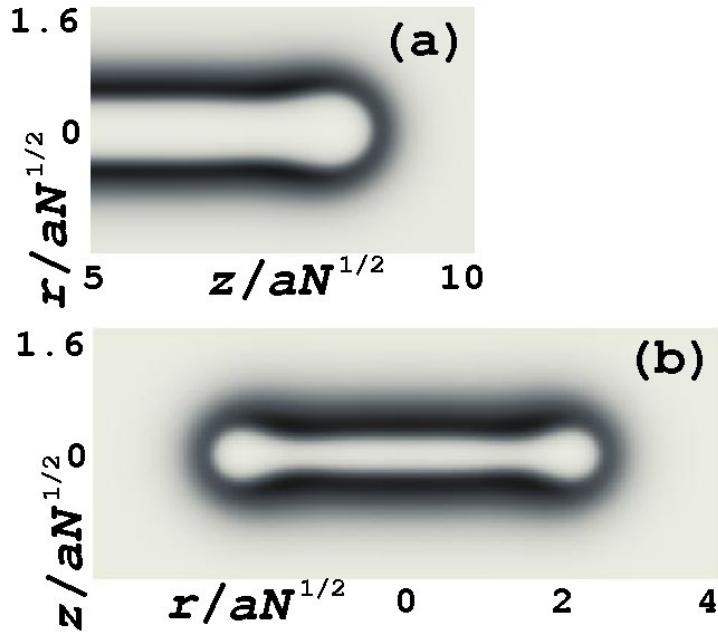

Figure 9: Density plots of A blocks in the system (due to either diblock species) for (a) rods (b) platelets. Dark areas show high volume fraction.

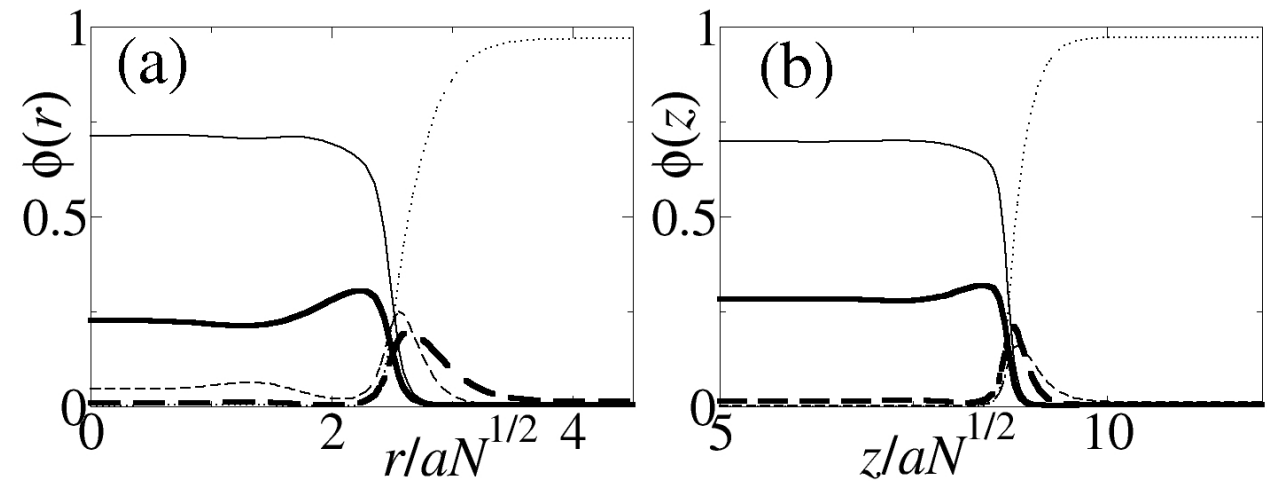

Figure 10: Cuts through the volume fraction profiles of the different AB diblock species for the (a) rod (b) platelet in Figure 8. A blocks are shown as dashed lines, B blocks as full lines and the A homopolymer as the dotted line. Thicker lines are used to highlight the sphere forming species. Panel (a) shows cuts through the cylindrical structure along the z direction at r = 0 while (b) shows cuts through the platelet structure along the r direction at z = 0.

Our calculations clearly demonstrate the coupling between micelle curvature and polymer composition and how this coupling helps to stabilize the edges of the wide range of finite structures observed for intermediate ratios of sphere formers. We note that Bates and co-workers have found undulating cylindrical structures in bimodal blends of PEO-PB block copolymers which contain similar negative curvature regions as in the bulbous rod ends.[17] Our calculations indicate that these undulating structures probably also arise from a coupling of curvature with polymer composition.

**Mixing at intermediate THF concentrations**

To study the effect of blending history on micelle morphology, in this section, we present results obtained by mixing the two copolymer solutions at intermediate stages during the self-assembly process ('intermediate mixing' protocol). The mixing procedure used is sketched in Figure 11. Firstly, two solutions of $PCL_{5k}PEO_{1k}$ and $PCL_{5k}PEO_{2k}$ at concentrations of 10 mg ml$^{-1}$ were prepared in THF. Both solutions were then diluted individually with water to the same extent (shown on the left side of Figure 11). At different stages of this dilution process, specifically at water contents (volume fraction) of 0, 20, 40, 60 and 72%, aliquots were taken and mixed in a ratio of 3 parts $PCL_{5k}PEO_{1k}$ solution to 1 part $PCL_{5k}PEO_{2k}$ solution (this ratio was chosen to match the pre- and post-mixing experiments described above). These mixtures were left to evolve for 1 hour at the respective water content of the mixed solutions before dilution to a final water content of 72%.

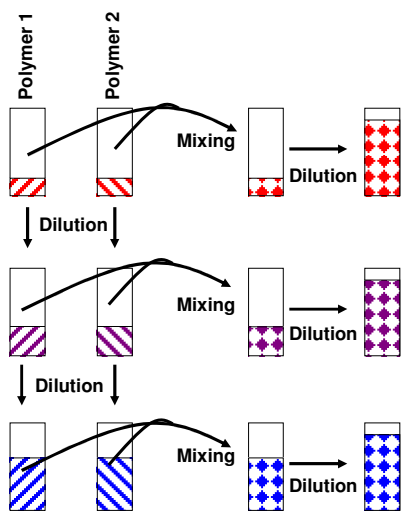

Figure 11: Scheme of the intermediate mixing protocol.

Visually, the $PCL_{5k}PEO_{2k}$ solutions remained clear during dilution. DLS and Cryo-TEM images in our previously published work[11] indicate that at 0 and 20% water content the $PCL_{5k}PEO_{2k}$ block copolymers are molecularly dissolved, while at 40% water content we have spherical micelles and short worms and finally at 60 and 72% water content we have spherical micelles. On the other hand, the $PCL_{5k}PEO_{1k}$ solutions are visually clear at 0% and 20% water content and turbid for 40% water content and above as can be seem from the turbidity trace in Figure 3 (open squares). For this system our previous cryo-TEM results showed a transition from molecularly dissolved block copolymers at 0% and 20% water content to worm-like micelles at 40% water content and finally vesicles at 60% and 72% water content.

For the mixed solutions (25% $PCL_{5k}PEO_{2k}$ or mixing ratio 3:1), the solutions mixed at 0% and 20% water content remained clear while the solutions mixed at 60% and 72% water content became slightly more turbid. However, the solution mixed at 40% water became completely clear instantly (<1s). This indicates that the pre-assembled worms present in the $PCL_{5k}PEO_{1k}$ solution are broken up in a fast process and suggests that inter-micellar exchange of block copolymer chains remains fast for water contents up to 40%. Thus for the solutions mixed at 40% and 60% water content, in addition to mixing different pre-formed structures (micelles, worms or vesicles), the block copolymers may still be sufficiently mobile so that significant fusion of the different structures can occur. If this is indeed the case, we may then reasonably expect these experiments to produce complex hybrid structures containing regions of markedly different local curvatures.

In Figure 12, we show representative cryo-TEM images of the solutions mixed at 20%, 40% and 60% water content, after the final dilution of these solutions to 72% water. For the solution mixed at 20% water (Figure 12a), only vesicles were found, similar to what was seen for the sample mixed in pure THF (pre-mixed). However for the solutions mixed at 40% (Figure 12b) and especially 60% water (Figures 12c, 13), intriguing new structures emerge which are highly reproducible. These include paddle-shape structures, 'horseshoes' (i.e., 'handle-free' variants of the paddles) and structures with the appearance of twisted membranes with rounded edges. However, while stable on timescales of many days, the metastability of these structures was apparent in experiments in which the solutions containing them were aged for 5 months, after which the zoo of odd shapes is replaced by a mixture of large vesicles that either have a single wall or up to three distinct walls which are at variable distances in the range of tens of nanometers apart (Figure 12d).

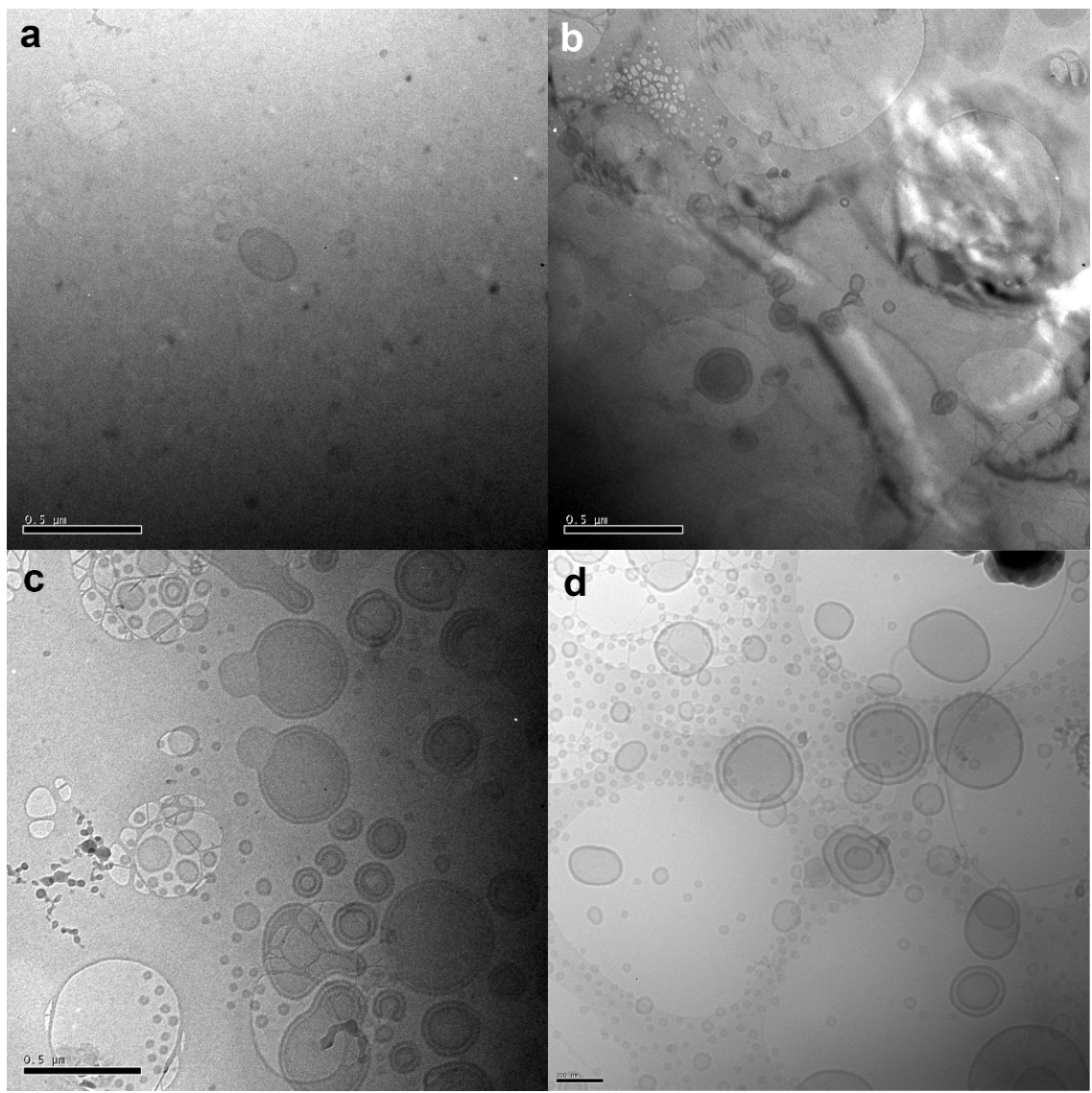

Figure 12: Cryo-TEM images obtained at 28% THF in water (72% water). PCL$_{5k}$PEO$_{1k}$ and PCL$_{5k}$PEO$_{2k}$ were mixed from partially self-assembled pure solutions at (a) 20% (b) 40% and (c) 60% water in a ratio of 3:1. (d) shows the solution imaged in (c) after ageing for 5 months at the imaging water concentration of 72%. All images have the same magnification; the scale bars are 500 nm (a-c) or 200 nm (d).

We note that similar horseshoe and paddle shaped aggregates have been observed in experimental studies using other chemistries of block copolymer[8,12,31] but there has been no detailed discussion regarding their structure or mechanism of formation. On the other hand, the methodology introduced here, where we combine theory with experiment, provides us with a deeper insight into their emergence. The constant small distance between many of the double edged structures in Figure 12c, 13 indicates that the two edges are strongly coupled and precludes the identification of these structures as double walled vesicles. Instead, the most likely structure suggested by the paddle forms in Figure 12c and on the right panel of Figure 13 is the 'nano-pouch' structure sketched in Figure 14. The different regions of density in the cryo-EM micrographs strongly suggest a smaller single lamellum and larger double region, the latter surrounded by an edge of controlled curvature. This is a structure that would plausibly emerge kinetically from the fusion of preexisting vesicles (made up of lamellar formers) and spherical micelles or short worms (made up of sphere formers). The emergence of a hybrid structure consisting of flat bilayers (arising from a flattening of the vesicle) surrounded by a near-toroidal edge presumably represents the systems attempt to accommodate the two species which have very different spontaneous curvatures. In particular our earlier theoretical analysis suggests that the torroidal edges are stabilized by an enrichment of sphere formers while the negatively curved regions inside the edges are stabilized by a depletion of sphere formers.

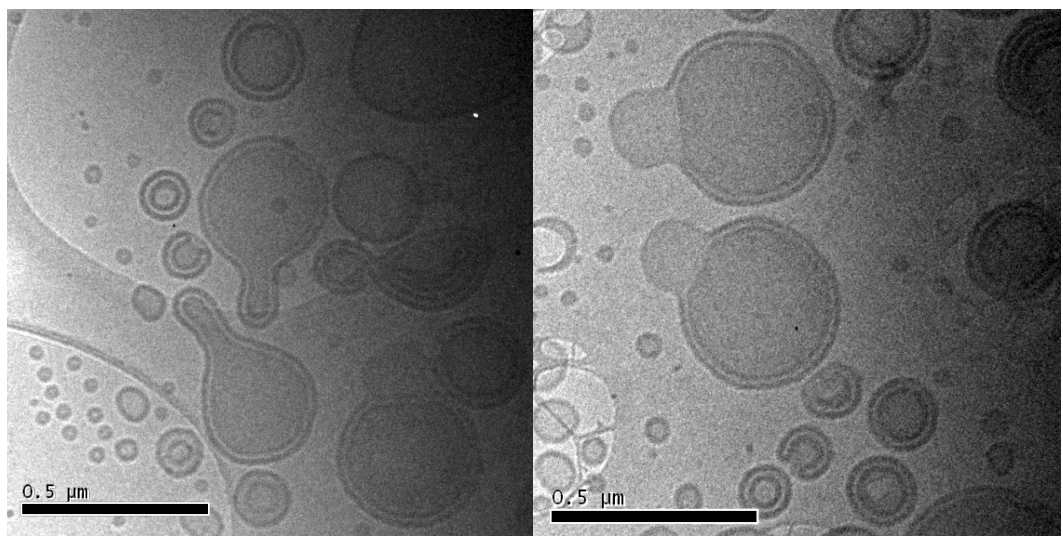

Figure 13: Close-up of some of the unusual shapes seen by cryo-TEM in samples mixed in partially assembled state (60% water) (see Figure c)).

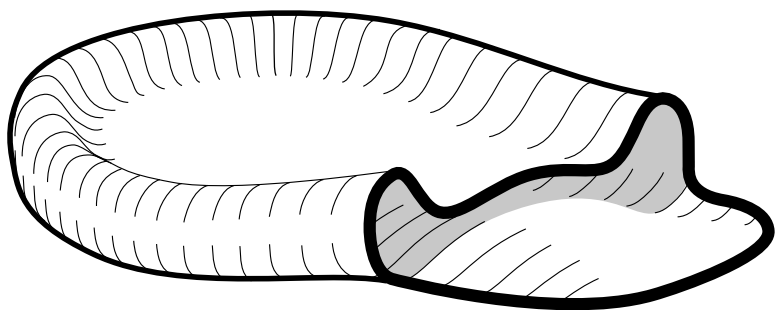

Figure 14: Sketch of the proposed morphology of the 'horseshoes' and 'paddles' shown in Figures 12c, 13

In a final step, we note that such a thermodynamically-driven flattening of the vesicle must increase its surface/volume ratio, and so burst the aggregate open at at least one point. There is no reason to assume that when this happens, equal areas of sheet should exist on either leaflet. The intermediate result of this is the formation of the 'handles' of Figure 13 (right panel) from the extra area of one leaflet. Although no isolated single lamellum was observed in our pre-mixed experiments, for the current volume fraction of sphere formers where vesicles are the equilibrium structures, our calculations in fact show that in the corresponding region of the free energy plot (0-10% region of Figure 7), the lamellum free energies are only slightly higher than that for vesicles, rings and rods. This may explain why the single lamellum can exist in these kinetically funnelled metastable structures. Subsequent transport of polymer within this structure will eventually equilibrate the material on each leaflet and allow the high-curvature edge to connect. The ubiquity of residual paddle structures even in the completed-edge structures of Figure 13 (left panel) support such a kinetic route.

**Conclusion**

We have studied the self-assembly of a binary mixture of polycaprolactone-polyethyleneoxide (PCL-PEO) block copolymers in solution where the two species on their own form vesicles and spherical micelles respectively in water and where self-assembly is triggered by changing the solvent from THF to water. We find that for this system, inter-micellar exchange of block copolymer molecules is extremely slow in water so that the resultant self-assembled structures are non-ergodic. This allowed us to control micelle morphology both thermodynamically and kinetically. In particular, we found that when the two species are fist molecularly dissolved in THF before mixing and self-assembly ('pre-mixing'), the micelle morphology depends strongly on the mixing ratio of the two species. Specifically, as we increase the proportion of sphere formers, the morphology changed from vesicles *via* an intermediate regime of small vesicles, 'bulbed' rods, Y-junctions and rings finally to spherical micelles. We note that in the intermediate regime (50% sphere formers), the micelle morphology is richer than the morphology of a monomodal solution with the same average volume fraction of PEO.

In other experiments where the two species are first (partially) self-assembled (by partial dilution of THF with water) before mixing and further self-assembly ('intermediate mixing'), novel metastable structures such as 'horseshoes', 'paddles' etc. emerge in cryo-TEM images. Careful analysis of the TEM micrographs suggest that these represent nanoscopic 'pouches' that arise from the fusion of relatively mobile, pre-formed aggregates which have very different curvatures.

The experimental observations above are corroborated by Self Consistent Field Theory (SCFT) calculations of the binary system. In particular, as we increased the proportion of sphere former, SCFT reproduced the same sequence of morphologies as that observed in the pre-mixing experiments and was also able to reproduce the shape for the bulbous rod ends. SCFT further showed that regions of strong positive curvature in the block copolymers aggregates, e.g., end caps in rods and edges in disks, are stabilized by a preferential enrichment with sphere formers while regions with negative curvature are stabilized by a depletion of sphere formers. Our calculations thus demonstrate quantitatively the coupling between local amphiphile composition and micelle curvature. This coupling also allows us to rationalize how the exotic structures containing regions of differential curvature that emerge from our intermediate-mixing experiments are stabilized.

In summary, both experiment and theory demonstrate that controlled blending of block copolymers is an effective design parameter for controlling the morphology of the self-assembled structures in block copolymer solutions and allows us to access a much richer range of nano-morphologies than is possible with single, monomodal block copolymer solutions.

**Acknowledgements:** This work was performed in project 264 of the micro and nanotechnology scheme part-funded by the UK Technology Strategy Board (formerly DTI). Unilever is thanked for permission to publish this work.

**For Table of Contents use only:**

Controlling the micellar morphology of binary PEO-PCL block copolymers in water-THF through controlled blending

Peter Schuetz[*], Martin J. Greenall, Julian Bent, Steve Furzeland, Derek Atkins, Michael F. Butler, Tom C.B. McLeish, D.Martin A.Buzza

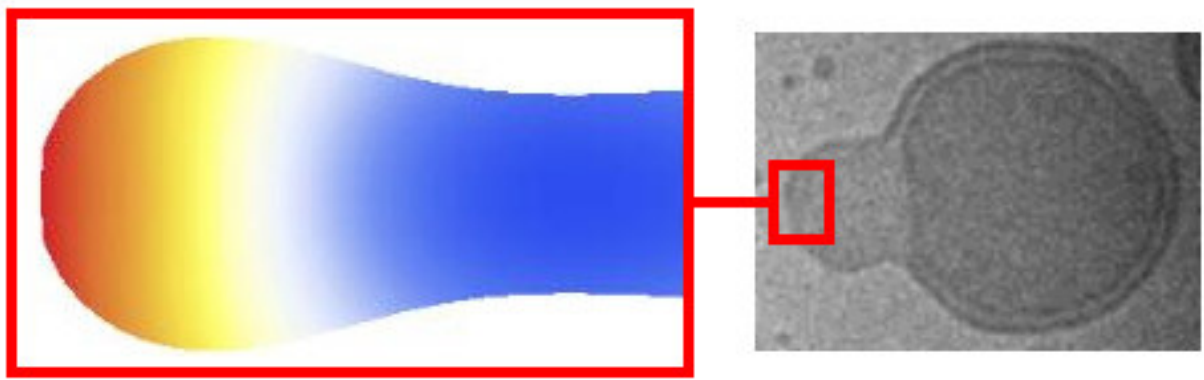

Or alternatively if a square image is required:

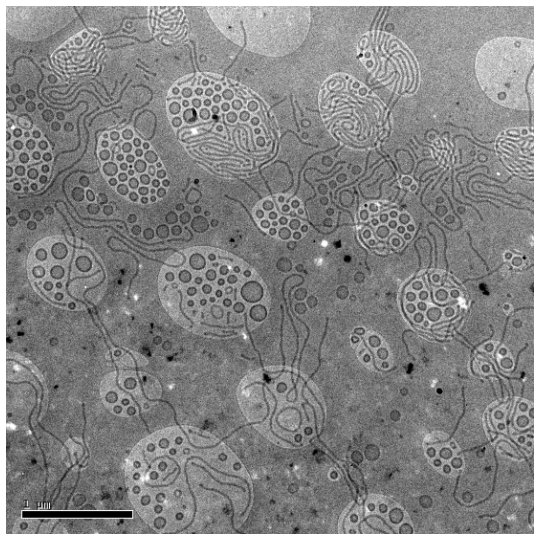